\documentclass[onecolumn,journal]{IEEEtran}
\IEEEoverridecommandlockouts
\usepackage{cite}
\usepackage{amsmath,amssymb,amsfonts}
\usepackage{graphicx}
\usepackage{textcomp}
\usepackage{algorithm}
\usepackage{xcolor}
\usepackage{float}
\usepackage{soul}
\usepackage{adjustbox}
\usepackage{booktabs}
\usepackage{url}  
\usepackage{verbatim}
\usepackage{colortbl}
\usepackage{setspace}

\newcommand{\comments}[1]{}
\usepackage{multirow}
\def\BibTeX{{\rm B\kern-.05em{\sc i\kern-.025em b}\kern-.08em
    T\kern-.1667em\lower.7ex\hbox{E}\kern-.125emX}}

\usepackage{algpseudocode}
\algrenewcommand\algorithmicrequire{\textbf{Input:}}
\algrenewcommand\algorithmicensure{\textbf{Output:}}

\title{Novel Fine-Tuned Attribute Weighted Naïve Bayes NLoS Classifier for UWB Positioning}

\begin{document}

\author{Fuhu Che$^{\star}$, 
        Qasim Zeeshan Ahmed$^{\star}$,   
        Fahd Ahmed Khan$^{\dagger}$,  and Faheem A. Khan$^{\star}$. 
\thanks{
$^{\star}$F. Che, Q. Z. Ahmed, and F. A. Khan is with the School of Computing and Engineering at the University of Huddersfield, United Kingdom. Email: \{fuhu.che, q.ahmed, f.khan\}@hud.ac.uk. 

$^{\dagger}$F. A. Khan is with the School of Electrical and Computer Engineering, University of Oklahoma, Tulsa, OK, USA. Email: fahd.khan@ou.edu. 

Part of this work was presented at the IEEE International Black Sea Conference on Communications and Networking 2022, Sofia, Bulgaria.}}
\maketitle

\begin{abstract}
In this paper, we propose a novel Fine-Tuned attribute Weighted Naïve Bayes (FT-WNB) classifier to identify the Line-of-Sight (LoS) and Non-Line-of-Sight (NLoS) for UltraWide Bandwidth (UWB) signals in an Indoor Positioning System (IPS). The FT-WNB classifier assigns each signal feature a specific weight and fine-tunes its probabilities to address the mismatch between the predicted and actual class. The performance of the FT-WNB classifier is compared with the state-of-the-art Machine Learning (ML) classifiers such as minimum Redundancy Maximum Relevance (mRMR)- $k$-Nearest Neighbour (KNN), Support Vector Machine (SVM), Decision Tree (DT), Naïve Bayes (NB), and Neural Network (NN). It is demonstrated that the proposed classifier outperforms other algorithms by achieving a high NLoS classification accuracy of $99.7\%$ with imbalanced data and $99.8\%$ with balanced data. The experimental results indicate that our proposed FT-WNB classifier significantly outperforms the existing state-of-the-art ML methods for LoS and NLoS signals in IPS in the considered scenario.
\end{abstract}

\begin{IEEEkeywords}
UWB, ML, WNB, mRMR, and IPS.
\end{IEEEkeywords}
\section{Introduction}
Mobile target tracking is extensively used in commercial, civil, industrial, and military applications, such as object tracking, autonomous vehicles, forklift control, military reconnaissance, etc~\cite{Zafari-2019, Ahmed-2014, Roy-2022}. Ultra-wideband (UWB) technology, utilizes exploratory signals with an extremely large spectrum, that provides significant advantages in tracking and positioning~\cite{Wymeersch-2012,Ahmed-2010,Ahmed-2011, Che-2022}. Beside this various alternative technologies have been extensively researched and employed in indoor scenarios such as Radio Frequency Identification (RFID), Bluetooth, WiFi fingerprinting, etc. Recently,~\cite{Zhang-2022, Roy-2022} has shown that WiFi with channel state information can be utilized for Indoor Positioning System (IPS) and have the same positioning cost and its accuracy is not inferior to UWB. Non-line-of-sight (NLoS) conditions are a major hindrance in achieving high positioning accuracy~\cite{Wang-2021a, Wang-2021b}. In NLoS, the direct signal path between the transmitters and the receivers is blocked by the obstacles causing estimation error~\cite{Che-2020, Che-2022}. To deal with the impact of NLoS, several signal-feature-based Machine Learning (ML) classification methods have been proposed~\cite{Jiang-2020, Stahlke-2022, Sang-2020, Wang-2021b}. 

Non-parametric regression technique based on Support Vector Machine (SVM) and Gaussian process to estimate the ranging error was proposed in~\cite{Wymeersch-2012}. It was shown that for signals under NLoS conditions, $28\%$ had a ranging error of less than one meter. Deep learning methods such as Convolutional Neural Network (CNN) and Long Short-Term Memory (LSTM) were employed in~\cite{Jiang-2020, Stahlke-2022}. The highest classification accuracy of $82.14\%$ was achieved by employing CNN and stacked LSTM together~\cite{Jiang-2020}. Other ML techniques such as SVM, Random Forest (RF), and Multi-Layer Perceptron (MLP) were applied in~\cite{Sang-2020} and an accuracy of $82.80\%$, $91.9\%$, and $91.20\%$ was reported for these techniques, respectively. A semi-supervised learning approach has also been adopted for NLoS identification. It was shown that NLoS identification accuracy could be improved to $94.7\%$ with the aid of semi-supervised SVM via leveraging unlabeled data~\cite{Wang-2021b}. An unsupervised ML approach based on Gaussian mixture models (EM-GMM) to identify the NLoS links from the unlabelled data was proposed in~\cite{Fan-2019}. \cite{Barralet-2019} employed the ranging and received signal strength (RSS) to identify the NLoS condition using LoS, NLoS soft, and NLoS hard classification. Furthermore, in~\cite{Vales-2020} the power delay profile was exploited to accelerate the training of NN-based classifiers. However, the dataset was perfectly balanced. Finally, in~\cite{Barral-2019a}, a different scenario was considered for testing with respect to the training for generalization of IPS. 

Feature-based ML methods were well-studied in the aforementioned works, however, none of these works considered the correlation between different features of the signals. Moreover, in an imbalanced dataset, where a limited amount of NLoS samples are present, it is difficult for existing algorithms to train a robust model~\cite{Jiang-2020}. To address these shortcomings, we propose a novel feature-based Fine-Tuned attribute Weighted Naïve Bayes (FT-WNB) classifier. To the best of the authors' knowledge, no work has proposed signal-feature weighting to enhance NLoS classification with an imbalanced dataset. The main contributions of this work are as follows:

\begin{itemize}
   \item A novel signal-feature-based FT-WNB algorithm for NLoS classification addresses class imbalance by assigning different weights to each feature and fine-tunes the classifier probabilities to improve classification accuracy.
  
  \item We conduct extensive experimentation to validate the superiority of the proposed FT-WNB classifier in comparison to the state-of-art algorithms. The performance is compared in terms of the confusion matrix, Receiver Operating Characteristics (ROC) curve, Area Under Curve (AUC), precision, recall, and classification accuracy.
  
  \item We show that the FT-WNB classification algorithm outperforms existing algorithms for different ratios of NLoS and LoS samples, even in the presence of imbalanced dataset.
\end{itemize}
\section{Fine-Tuned Weighted Naïve Bayes (FT-WNB) Classification Algorithm}~\label{sec:Section-2}
\vspace{-1cm}
\subsection{UWB Range Theory}
UWB transmission and detection algorithms can be found in more details in~\cite{Ahmed-2008, Ahmed-2008a}. Time-of-Arrival (ToA) is used to calculate the estimated distance $d$ between the tag and the anchor~\cite{Stahlke-2022},
\begin{equation}\label{eq-1}
d= 
\begin{cases}
c \times \mathrm{ToA} + \epsilon,                     & \ \   \mathrm{LoS}\\
c \times \mathrm{ToA} + \epsilon + b,    & \ \  \mathrm{NLoS}\\
\end{cases}
\end{equation}
where, $c$ is the speed of light in $\mathrm{m/s}$,  $\mathrm{ToA}$ is the propagation time taken from the tag to the anchor, $\epsilon$ represents the measurement noise and $b$ is the NLoS measurement error caused by the blockages and obstacles. 

\subsection{Weighted Naïve Bayes}

Naïve Bayes (NB) classifies the data with a corresponding maximum posterior probability. As features are assumed to be conditionally independent, the predicted class $\hat{l}^k$ at the $k$-th instance of data can be determined as 
\begin{eqnarray}\label{eq-2}
\hat{l}^k= \mathop{\arg\max}_{l} P(l)\prod_{i=1}^{I}P(X_i^k|l),
\end{eqnarray}
where $P(X_i)$ is the probability of the $i$-th attribute, $P(l)$ is the apriori probability of class $l$ and $I$ is the number of attributes. The underlying principle of WNB algorithm is that some attributes are of more importance than others~\cite{Ruan-2020}. This importance is specified by weighting the attributes and the predicted class, $\hat{l}^k$, in this case, is determined as 
\begin{eqnarray}\label{eq-3}
\hat{l}^k= \mathop{\arg\max}_{l} P(l)\prod_{i=1}^{I}P({X_{i}^k| l})^{w(i)},
\end{eqnarray}
where $w(i)$ is the weight associated with the $i$-th attribute.

    .

%
\begin{figure}[!t]
    \centering
    \includegraphics[width=1.0\linewidth]{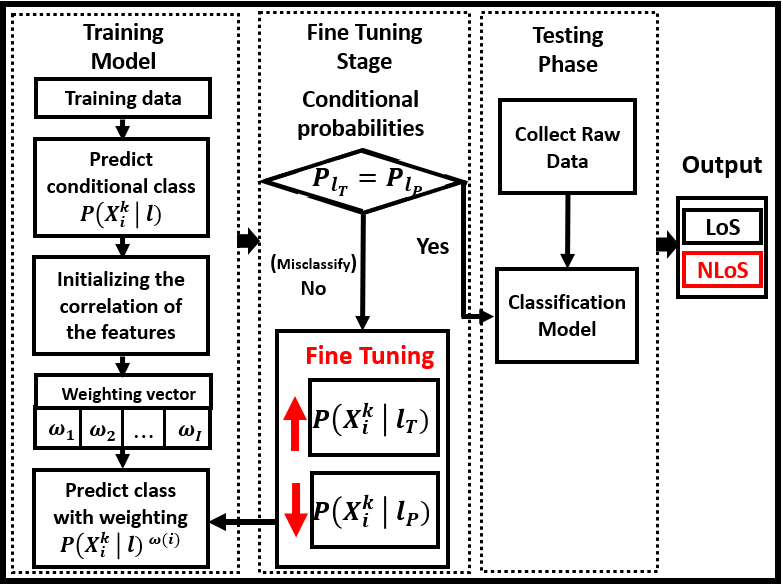}
    \caption{Block diagram of Fine-tuned WNB algorithm.}
    \label{fig-1}
\end{figure}
\begin{figure}[!t]
    \centering
    \includegraphics[width=1.0\linewidth]{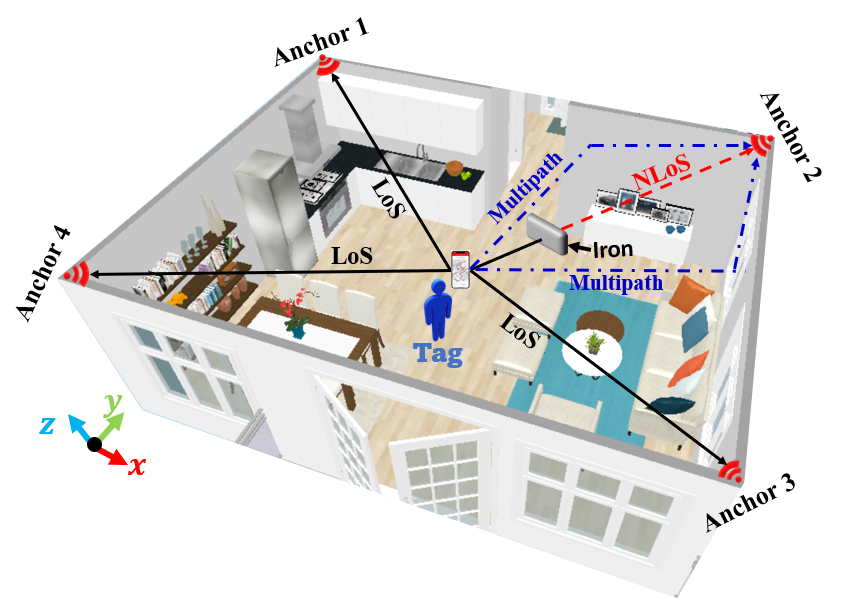}
    \caption{Experimental data collected in the studio.}
    \label{fig-2}
\end{figure}
%
\subsection{FT-WNB Algorithm}
%
\begin{figure*}[!t]
    \centering
    \includegraphics[height=0.4\linewidth, width=0.55\linewidth]{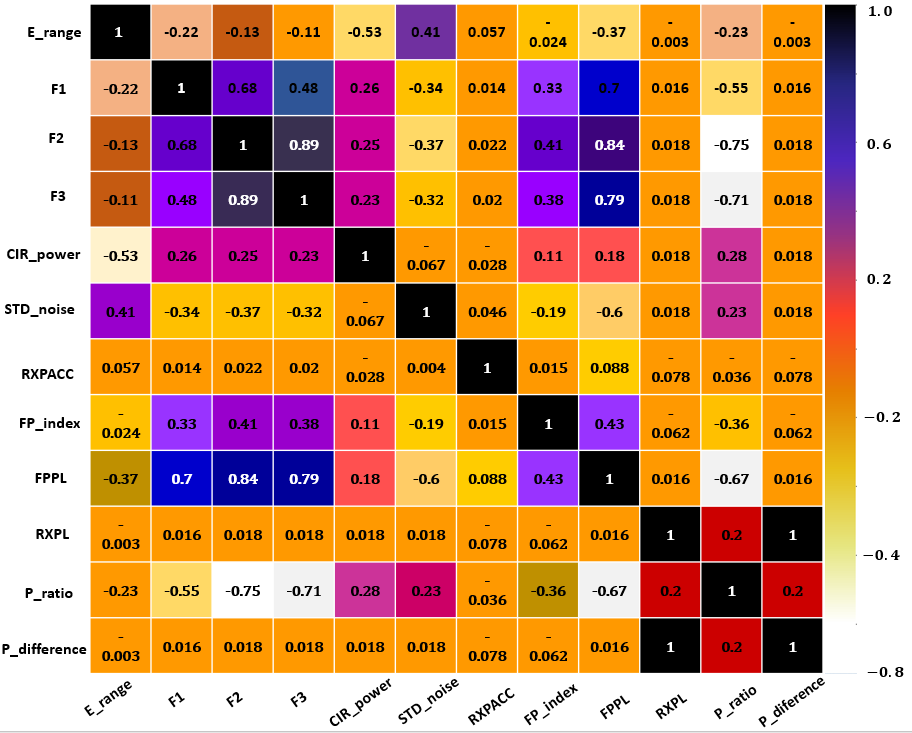}
    \caption{Signal-features correlation matrix.}
    \label{fig-3}
\end{figure*}
The overall framework of the FT-WNB algorithm is described in Fig.~\ref{fig-1}. We first initialize the conditional probabilities as defined in (\ref{eq-3}) by applying the attribute weights to predict the class label. In the training phase, if the training instance is classified incorrectly, the predicted class $l_{P}$ has a higher posterior probability than its true class $l_{T}$. Therefore, the conditional probabilities of signal features, $P(X_i^k | l_{T})$ and $P(X_i^k | l_{P})$ can be fine tuned in the next iteration as 
\begin{eqnarray}\label{eq-4-5}
P_{t+1}(X_i^k | l_{T}) &=& P_{t}(X_i^k | l_{T}) + \xi_{t+1}(X_i^k , l_{T}),\\
P_{t+1}(X_i^k | l_{P}) &=& P_{t}(X_i^k | l_{P}) - \xi_{t+1}(X_i^k , l_{P}),
\end{eqnarray}
where $t$ represents the number of iteration and $\xi_{t+1}(\cdot)$ are the updated step size calculated as 
\begin{eqnarray}\label{eq-6-7}
\!\!\!\!\xi_{t+1}(X_i^k,l_{T})\!\!\!\!\!\!&=&\!\!\!\!\!\!\beta\!\!\left(\alpha\!\! \left(\max P_t(X_i^k |l_{T})\right)\!\!-\!\!P(X_i^k | l_{T})\right)\!\!e(l_T,l_P),\\
\!\!\!\!\xi_{t+1}(X_i^k,l_{P})\!\!\!\!\!\!&=&\!\!\!\!\!\!\beta\!\!\left(\alpha P(X_i^k| l_{P})\!\!-\!\!\min\left(P_t(X_i^k| l_{P})\right)\right)\!\!e(l_T,l_P),
\end{eqnarray}
where $\alpha, \beta \in (0,1)$ controls the size of the updated step for $P_{t}(X_i^k | l_{T})$ and  $P_{t}(X_i^k | l_{P})$, respectively. $\max (P_t(X_i^k |l_{T}))$ and $\min(P_t(X_i^k |l_{P}))$ are the maximum and the minimum value for $P_{t}(X_i^k | l_{T})$ and $P_{t}(X_i^k | l_{P})$, respectively. The error term, $e(l_T,l_P)$, is calculated as 
\begin{eqnarray}\label{eq-10}
e(l_T,l_P) = |P(l_T) - P(l_P)|.
\end{eqnarray}
Finally, the steps involved in the training and testing phase are summarised in Algorithms $1$ and $2$, respectively.

\begin{algorithm}[t]
\caption{: Training Phase}
\begin{algorithmic}
\Require Training dataset consisting of signal features.
\Ensure The conditional probability vector\\
\textbf{Phase 1: Initializing conditional probabilities}\\
$\square \quad \textbf{\textit{Step 1:}}$  \For {each $X_i^k$}
				\State calculate $\hat{l}^k$ by (2)
			\EndFor
			\For {each weighted $X_i^k$}
				\State calculate $\hat{l}^k$ by (3)
			\EndFor ~~~Estimate the conditional probability vector\\
\textbf{Phase 2: Fine tuning conditional probabilities} \\
$\square \quad \textbf{\textit{Step 2:}}$
        \If{$l_{T}\neq l_{P}$}
            \State Calculate the error by (8)
            \State Calculate the $\max(P_t(X_i^k |l_{T}))$ and $\min(P_t(X_i^k|l_{P}))$
			\For {each weighted figure $X_i^k$}
				\State calculate $\xi_{t+1}(X_i^k , l_{T})$ by (6)
				\State update $P_{t+1}(X_i^k|l_{T})$ by (4)
				\State calculate $\xi_{t+1}(X_i^k , l_{P})$ by (7)
				\State update $P_{t+1}(X_i^k|l_{P})$ by (5)
			\EndFor            
        \EndIf
                \If{$l_{T} = l_{P}$}
        \EndIf
    \State \textbf{return} Classification model
\end{algorithmic}
\end{algorithm}
\begin{algorithm}[!tb]
\caption{: Testing Phase}\label{algorithm-2}
\begin{algorithmic}
\Require Test dataset with a mixture of LoS and NLoS signals
    \Ensure  Predicted class label of $X_i^k$\\
    \textbf{Algorithm:}
    \begin{enumerate}
      \item Estimate the class membership probability $P(X_i^k|l)$.
    \item Predict the class label $\hat{l}^k$ of each class $X_i^k$. 
    \item \textbf{Return} $\hat{l}^k$. 
    \end{enumerate}
    \end{algorithmic}
    \end{algorithm}
%
\section{Experimental Setup and Data Collection}~\label{sec:Section-3}
The required dataset was collected in a studio environment ($4.8m \times 3.5m$). The floorplan of the infrastructure is shown in Fig.~\ref{fig-2}. Decawave\textcircled{R} MDEK-$1001$ UWB kits, based on DW-1000 chip, with configuration as mentioned in~\cite{Decawave}, are used to generate the dataset. During data collection for the LoS scenario, there is no obstacle between the anchor and tag. For the NLoS scenario, an iron sheet is placed between the anchors and the tag. The plates are placed very close to the tag, generally less than a meter away. In this case, the first path of the signal transmission is either completely blocked or distorted through multiple reflections. As a result of this, the estimated range will be biased due to the signal propagation delay. From the collected dataset, we randomly select $100$ NLoS and $1000$ LoS signals to generate data imbalance. We consider two more features in addition to the $10$ features of~\cite{Che-2022a}: i) the index of the detected first-path ($FP\_index$) and ii) the power ratio between the estimated received power and first-path power. The features and the correlation between them are illustrated in Fig.~\ref{fig-3}. As expected, the features are clearly not equally important nor independent. It can be observed that the features amplitude of the first path ($F1$), the amplitude of the second path ($F2$), and  the amplitude of the third path ($F3$), as well as first path power level ($FPPL$), have a relatively higher correlation than others. 
\begin{figure}[!t]
     \centering
      \vspace{-0.5cm}
 \includegraphics[width=0.95\linewidth]{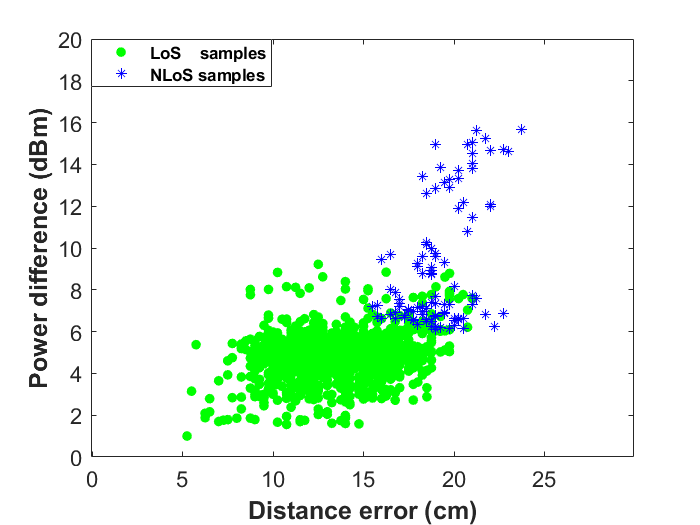}
        \caption{Samples of Error vs. Power Difference}~\label{fig-4}
 \end{figure}
 \vspace{-0.5cm}
\section{Performance Evaluation}~\label{sec:Section-4}
Visualization of our samples is shown in Fig.~\ref{fig-4} where the power difference versus the distance error is plotted. There are $1000$ LoS samples and $100$ NLoS signals, respectively. In this figure, the green samples represent the LoS signals and the blue color represents the NLoS signals. From the figure, it can be observed that few NLoS samples can be easily differentiated from LoS samples. While a large number of samples for NLoS cannot be easily differentiated from LoS samples. Therefore, for accurate classification, an appropriate algorithm is required.

\begin{figure}[!t]
    \centering
    \includegraphics[width=0.9\linewidth]{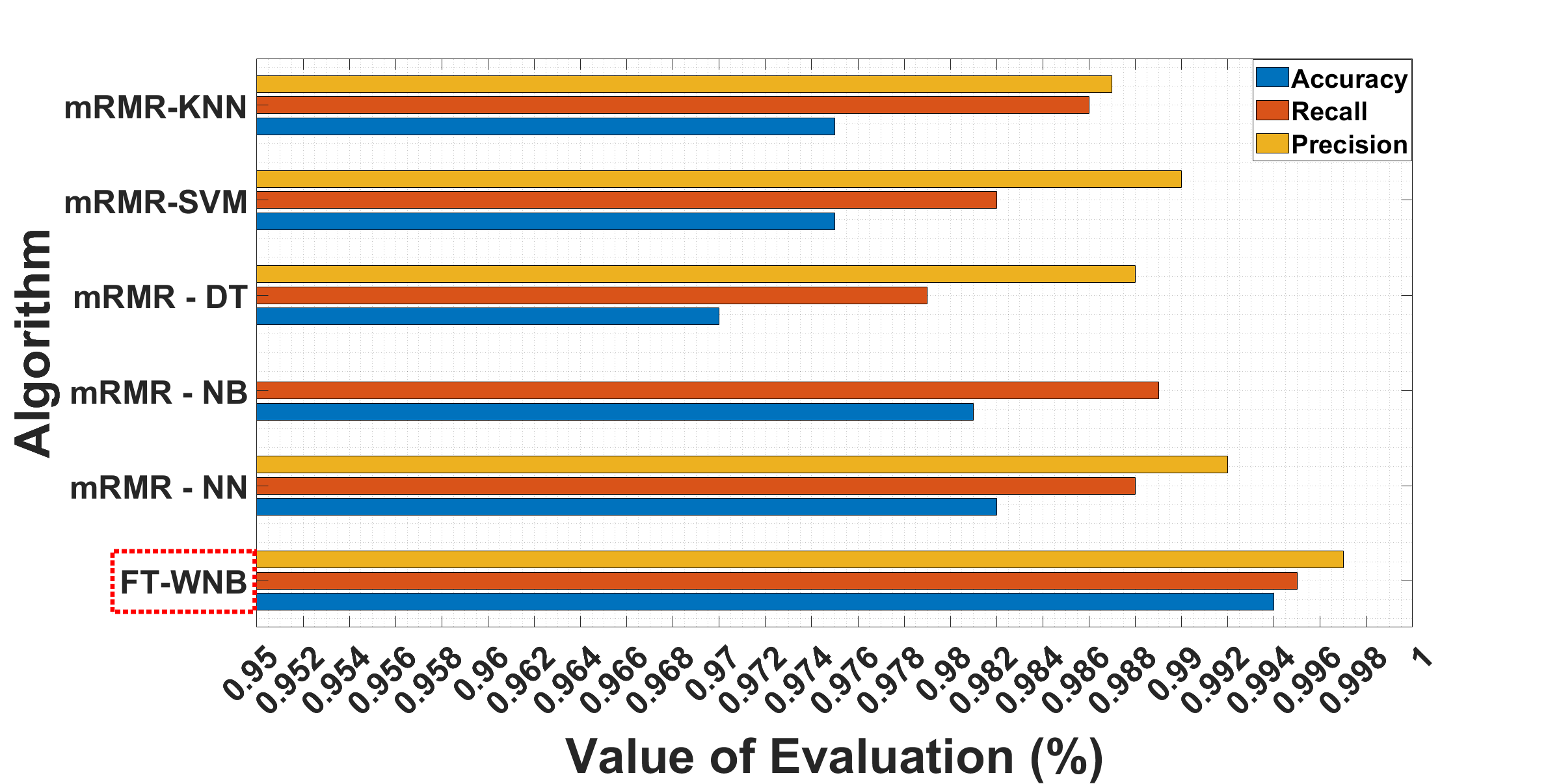}
    \caption{Performance comparison for the proposed FT-WNB algorithm with state-of-the-art ML algorithms.}
    \label{fig-5}
\end{figure}
\begin{table}
\caption{Running time and Confusion matrix of the proposed FT-WNB algorithm and state-of-the-art ML algorithms } 
\centering
\resizebox{1\columnwidth}{!}{%
\begin{tabular}{|l|c|c|c|c|c|c|c|c|}
\hline
& \textbf{Running}  & \textbf{LoS} & \textbf{NLoS}  & & & & & \\
\textbf{Algorithms} & \textbf{Time (s)} & \textbf{CR} & \textbf{CR}     & \textbf{TP} & \textbf{FN} & \textbf{FP} & \textbf{TN} & \textbf{Accuracy($\%$)}\\
\hline
\textbf{mRMR-KNN}    &  0.049    & 98.6\%    & 87\%   & 986   & 14   &13   &87   &97.5 \\
\textbf{mRMR-SVM}    &  0.092    & 98.2\%    & 90\%   & 982   & 18   &10   &90    &97.5\\
\textbf{mRMR - DT}   &  0.104    & 98.4\%    & 89\%   & 984   & 16   &11   &89    &97.5\\
\textbf{mRMR - NB}   &  0.048    & 98.9\%    & 90\%   & 989   & 11   &10   &90    &98.1\\
\textbf{mRMR - NN}   &  0.061    & 98.8\%    & 92\%   & 988   & 12   &8    &92    &98.2\\
\textbf{FT-WNB}      &  0.041    & 99.5\%    & 98\%   & 995   & 5    &2    &98    &\textcolor{red}{99.4}\\
\bottomrule
\end{tabular}%
}
\label{tab-1}
\end{table}

The performance of the proposed FT-WNB classifier is compared with ML classifiers in Fig.~\ref{fig-5}. For these classifier, a subset of features is selected based on minimum redundancy maximum relevance (mRMR) feature selection. mRMR is a feature selection approach that tends to select features with a high correlation with the class and a low correlation between themselves to maximize relevance toward the features and simultaneously to minimize the redundancy among features~\cite{Samat-2021}. The running time, confusion matrix and the correct rate (CR) is summarised in Table~\ref{tab-1}. The confusion matrix depicts metrics, True Positive (TP), False Positive (FP), False Negative (FN) and True Negative (TN), respectively. In order to compare the classification performance, we plot precision, recall and classification accuracy in Fig.~\ref{fig-5}. It can be observed that the best performing algorithm among the existing ML algorithms is the mRMR-NN, which has a precision of approximately $99.2\%$, recall around $98.8\%$ and overall classification accuracy is $98.2\%$. mRMR-NB follows closely and has similar performance. However, it is evident from Fig.~\ref{fig-5}, that the proposed FT-WNB algorithm outperforms all the state-of-the-art ML algorithms, i.e., mRMR-KNN, -SVM, -DT, -NB and -NN by achieving a precision of $99.7\%$, recall of $99.5\%$ and an overall classification accuracy of $99.4\%$. 
From Table~\ref{tab-1} it can be observed that the proposed FT-WNB algorithm, TN$=98$ which indicates that $98\%$ NLoS samples are correctly classified and only $2\%$ NLoS samples are incorrectly classified. The classification performance is significantly better than the second best case which is $92\%$ achieved by mRMR-NN algorithm. TP$=995$, refers to LoS resulting in a correct rate of $99.5\%$ which means $5$ samples out of $1000$ samples were inaccurately classified. Again, it is $1.3\%$ higher than the second-best case. The average running time of FT-WNB is $0.041$s which is better than other considered algorithms where the process of mRMR takes significant time. To sum up, in comparison to mRMR-KNN, mRMR-SVM, mRMR-DT, mRMR-NB, and mRMR-NN ML algorithms, it can be observed that the FT-WNB algorithm performs better for all performance metrics of the confusion matrix.

\begin{figure}[!t]
    \centering
    \includegraphics[width=0.9\linewidth]{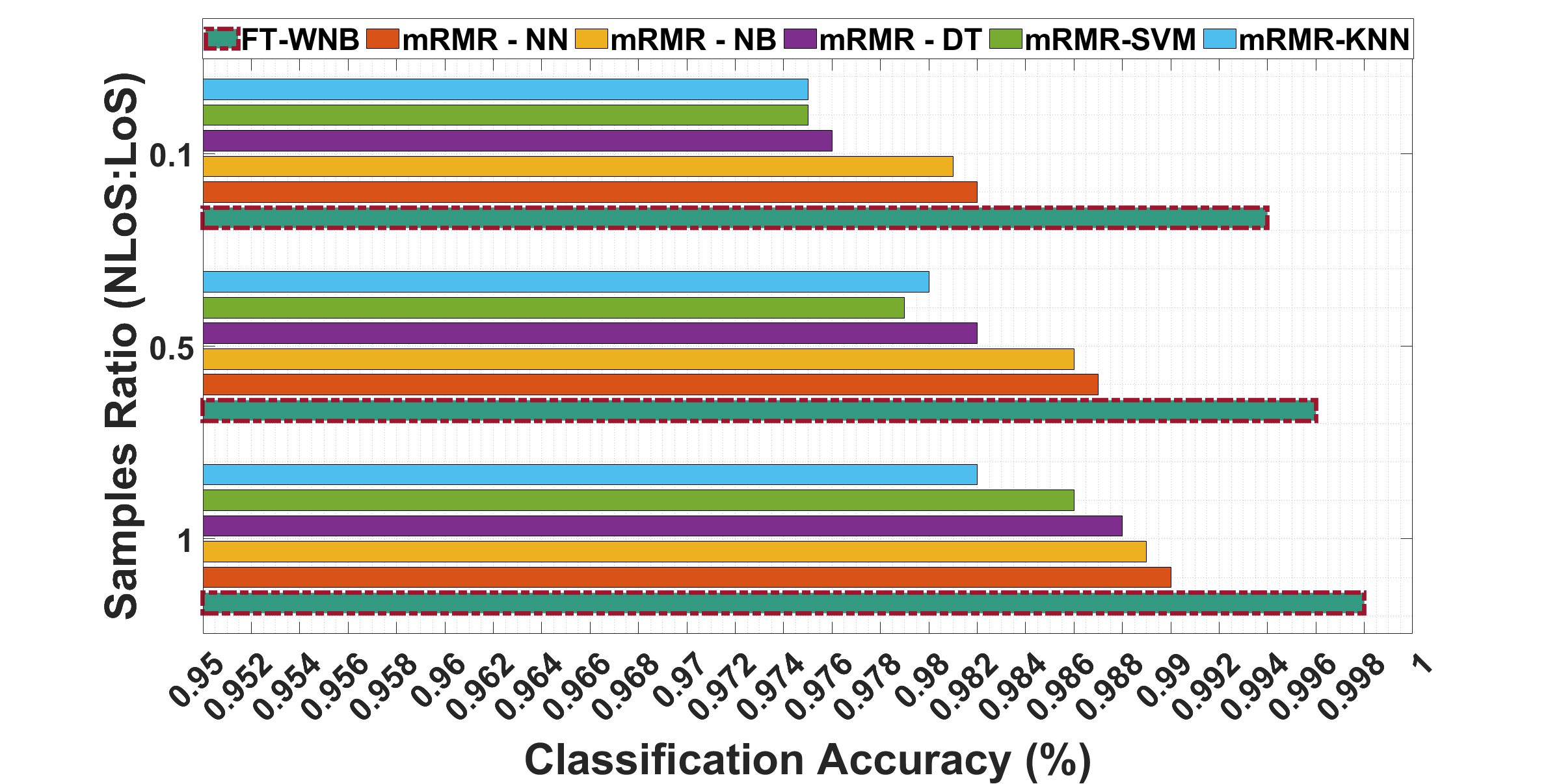}
    \caption{Classification accuracy of the FT-WNB algorithm under imbalanced datasets having varying ratios of data.}
    \label{fig-6}
\end{figure}

\begin{figure}[!t]
    \centering
    \includegraphics[width=0.9\linewidth]{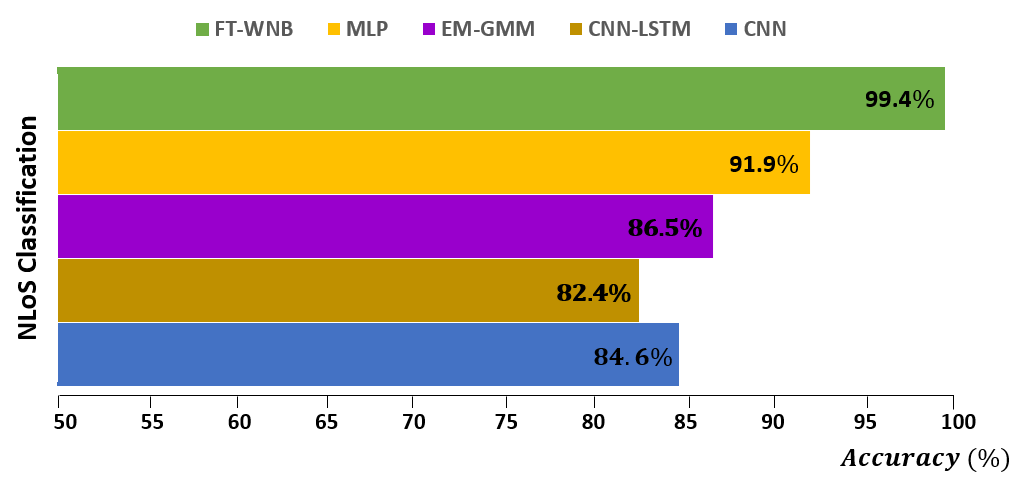}
    \caption{Comparisons for the existing UWB position methods.}
    \label{fig-7}
\end{figure}
Fig.~\ref{fig-6} shows the overall classification accuracy in the case of an imbalanced dataset. The ratio (NLoS: LoS) was set to $0.1$, $0.5$, and $1.0$, respectively. It can be observed from Fig.~\ref{fig-6} that as the ratio increases, the results of all algorithms improve because the model is trained better with equal number of LoS and NLoS samples. However, it is noticeable that the FT-WNB algorithm always maintains a higher classification accuracy for all considered ratios. This shows that the proposed FT-WNB classifier is highly robust and can assign appropriate weights to the desired features, guaranteeing better NLoS identification as compared to existing ML algorithms.

We also compares the proposed algorithm with the existing UWB NLoS identification methods as mentioned in the introduction section. The results in terms of accuracy are shown in Fig.~\ref{fig-7}. It can be clearly observed that the proposed FT-WNB algorithm performs better than MLP, EM-GMM, MLP, CNN-LSTM, and CNN methods which are some of the existing positioning methods in UWB IPS.

\begin{figure}[!t]
    \centering
    \includegraphics[width=0.8\linewidth]{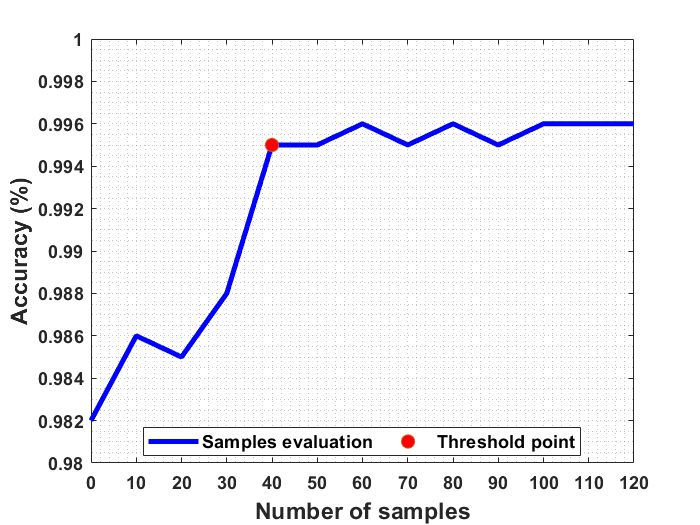}
    \caption{Impact of number of samples evaluation}
    \label{fig-8}
\end{figure}

For the fine-tuning step, the impact of a number of selected samples is shown in Fig.~\ref{fig-8}. It can be observed that the accuracy of the algorithm improves as the number of samples increase and the proposed model benefits from the fine-tuning step. The classification accuracy improves from $98.2\%$ to $98.6\%$ by collecting the first $10$ samples and then further increases to $99.5\%$ by just collecting the first $40$ samples. However, there is no significant gain in accuracy beyond that point. Therefore, in our simulations, the number of samples required for fine-tuning is fixed at $40$.

\begin{table}[!t]
\centering
\caption{Comparison with ~\cite{Barralet-2019}.}
\begin{tabular}{|l|l|l|l|l|}
\hline
\centering
\textbf{Algorithms}&\textbf{Processing}&\textbf{$\%$ LoS} &\textbf{$\%$ NLoS}&\textbf{$\%$ Total}\\
&\textbf{Time (s)}&\textbf{Accuracy}&\textbf{Accuracy}&\textbf{Accuracy}\\
\hline
 Range \& RSS      &  0.0371     &      97.4    &            91          &     96.8   \\
\hline
All features    & 0.041       &      99.5       &         98        &    99.4   \\
\hline 
Difference & 3.9~ms&2.1~($\%$)&7~($\%$)&2.6~($\%$)\\
\hline
\end{tabular}
\label{tab:check-1}
\end{table}
There is a tradeoff between the collected features and accuracy. In this paper, our focus was to collect a maximum number of features that yield the highest accuracy. In Table~\ref{tab:check-1}, we present the comparison of the processing time considering only two features (Range and RSS ~\cite{Barralet-2019}) and all the $12$ gathered features. It can be observed that a difference of $3.9$ms leads to an improved accuracy of $99.4\%$ as compared to $96.8\%$. However, if the application does not require high accuracy, we could reduce the number of collected features and improve the processing time of the algorithm.

Our algorithm has been tested and trained in various scenarios as adopted in~\cite{Barral-2019a}. It can be observed from Table~\ref{tab:Table-3} that the proposed WNB has a classification accuracy of over $98\%$, even after being trained in a studio and tested in a room measuring $(4.8m \times 5.4m)$. This demonstrates the versatility of FT-WNB and its ability to perform in different environments, regardless of where it was trained.
\begin{table}[!t]
\centering
\caption{Performance of FT-WNB in a different environment.}
\begin{tabular}{|l|l|l|l|l|l|l|}
\hline
\centering
\textbf{Training}&\textbf{Testing} &\textbf{TP} &\textbf{FP}&\textbf{FN}&\textbf{TN}& \textbf{Accuracy}   \\
&\textbf{Scenario}&\textbf{Scenario}&&&&($\%$)\\
\hline
 Studio   &Studio   & 995    & 5    &2   & 98    & 99.4 \\
\hline
 Studio   &Room     & 990   & 10    &5   & 95    & 98.6  \\
\hline
\end{tabular}
\label{tab:Table-3}
\end{table}
%
\section{Conclusion}~\label{sec:Section-5}

In this work, we propose the FT-WNB algorithm which is a feature-based NLoS classification method to improve classification accuracy by automatically weighting different signal features, for imbalanced LoS and NLoS samples. The performance of the FT-WNB classifier is compared with state-of-the-art machine learning algorithms using simulation. It is shown that the proposed FT-WNB algorithm performed better in terms of confusion matrix, ROC, and AUC. In addition, the proposed FT-WNB algorithm adapts better to the different ratios of LoS and NLoS signal data. Our findings demonstrate that the proposed FT-WNB classifier efficiently alleviates the influence of the signal features in an imbalanced dataset in a harsh mixed LoS and NLoS scenario.

\end{document}